\documentclass[pra,floatfix,twocolumn,superscriptaddress,showpacs,preprintnumbers,longbibliography,nofootinbib]{revtex4-1}

\usepackage[a4paper,left=1cm,right=1cm,top=2.5cm,bottom=2.5cm]{geometry}
\usepackage[utf8]{inputenc}
\usepackage[usenames,dvipsnames]{xcolor}
\usepackage[colorlinks=true,citecolor=Blue,urlcolor=Blue,linkcolor=Blue]{hyperref}
\usepackage[cm]{fullpage}
\usepackage{graphicx}
\usepackage[english]{babel}
\usepackage{amsmath}
\usepackage{amssymb}
\usepackage{cancel}
\usepackage{natbib}
\usepackage{comment}
\usepackage{color}
\usepackage[normalem]{ulem}

\usepackage{physics} 
\usepackage{bbold}

\newcommand{\T}[1]{\text{#1}}

\begin{document}

\title{Controlling Markovianity with Chiral Giant Atoms}

\author{Federico Roccati}
\affiliation{Department of Physics and Materials Science, University of Luxembourg, L-1511 Luxembourg}
\affiliation{Department of Physics, Columbia University, New York, New York 10027, USA}

\author{Dario Cilluffo}
\affiliation{Institut f\"ur Theoretische Physik and IQST, Albert-Einstein-Allee 11, Universit\"at Ulm, 89069 Ulm, Germany}


\begin{abstract}

    Giant artificial atoms are promising and flexible building blocks for the implementation of analog quantum simulators.
    They are realized via a multi-local pattern of couplings of two-level systems to a waveguide, or to a two-dimensional photonic bath.
    A hallmark of giant-atom physics is their \textit{non-Markovian} character in the form of self-coherent feedback, leading, \textit{e.g.}, to non-exponential atomic decay. 
    The timescale of their non-Markovianity is essentially given  by the time delay proportional to the distance between the various coupling points.
    In parallel, with the state-of-the-art experimental setups, it is possible to engineer complex phases in the atom-light couplings.
    Such phases simulate an artificial magnetic field, yielding a \textit{chiral} behavior of the atom-light system.
    Here, we report a surprising connection between these two seemingly unrelated features of giant atoms, showing that the chirality of a giant atom controls its Markovianity.
    In particular, by adjusting the couplings' phases, a giant atom can, counterintuitively, enter an exact Markovian regime irrespectively 
    of any inherent time delay.
    We illustrate this mechanism as an interference process and via a collision model picture.
    Our findings significantly advance the understanding of giant atom physics, 
    and open new avenues for the control of quantum nanophotonic networks. 
 
\end{abstract}

\maketitle

\begin{figure}
	\centering
	\includegraphics[width=\columnwidth]{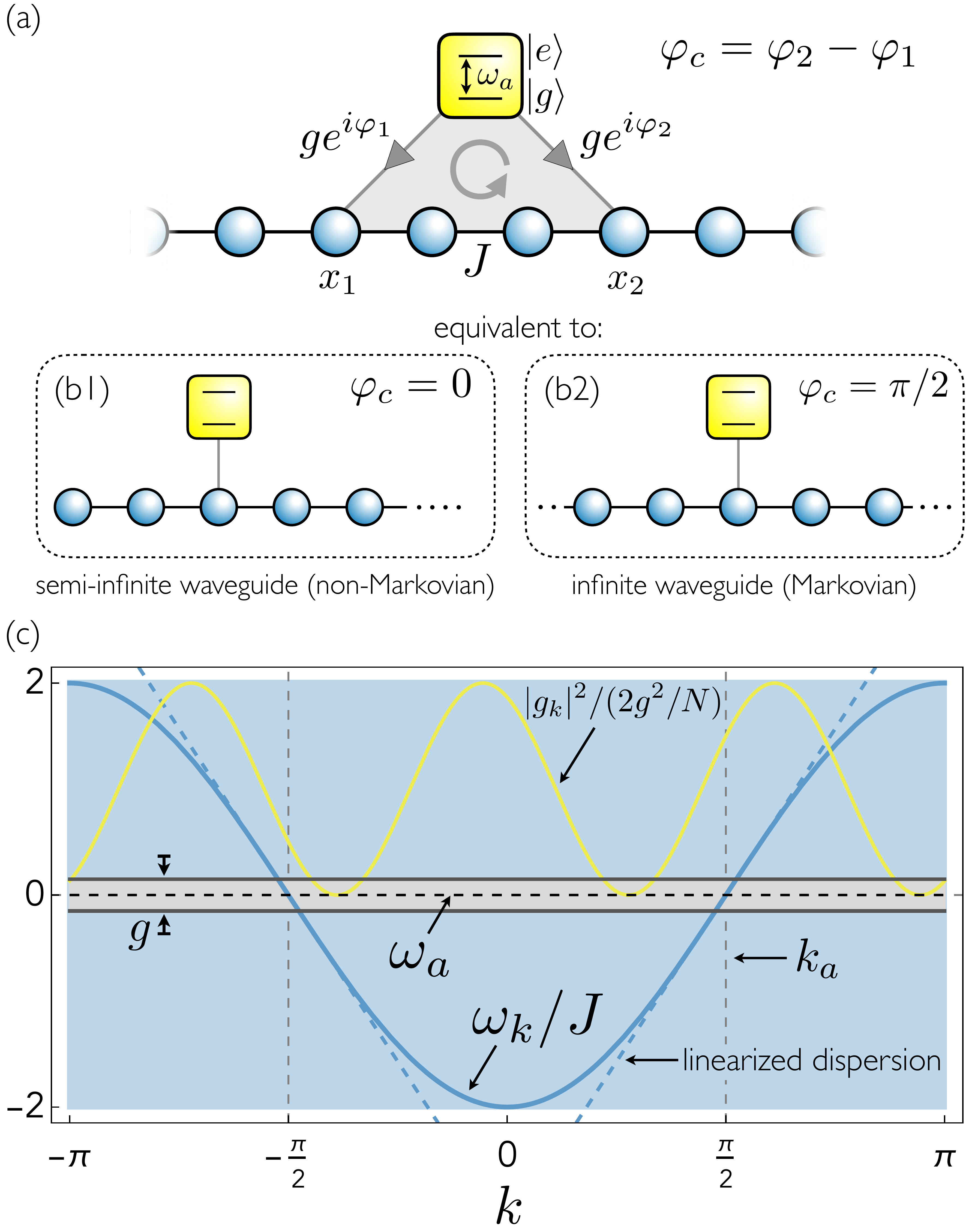}
	\caption{
		{\textit{Setup.} 
			(a) Chiral giant atom coupled at the points $x_{1,2}$ to a 1D bidirectional waveguide. The couplings are generally complex with phases $\varphi_{1,2}$, cf.~Eq.~\eqref{eq:int_ham}. When $\varphi_c=\pi/2$ [$\varphi_c=0$], as in (b2) [(b1)], the system is equivalent to a small atom (local coupling) in a [semi-]infinite waveguide and the dynamics is [non-]Markovian.
			(c) Dispersion law of the waveguide (solid blue), which in weak coupling (gray stripe) can be linearized (dashed blue). The atomic frequency $\omega_a$ is resonant with the middle of the band, corresponding to the momentum $k_a=\pi/2$. In yellow we show an instance of  a chiral  (indeed, $|g_{k_a}|>|g_{-k_a}|$) atom-waveguide coupling with $x_2-x_1=3$ and $\varphi_c=\pi/6$. }
	} 
	\label{FigAnalogy}
\end{figure}

The growing demand to process quantum information for computational purposes underscores the increasing importance of developing scalable quantum networks.
These networks consist of spatially distributed nodes interconnected by communication lines.
Consequently, investigating the realm where memory effects and quantum feedback are not negligible becomes increasingly crucial in addressing the challenge of quantum computation~\cite{LorenzoPRA2013,RamosPRA2016,HaasePRL2018,FangNJP2018,MilzPRX2020,WhiteNatComun2020,tserkis2023information}.
A notable instance includes multi-local or \textit{giant} atoms \cite{KannanNature2020,Kockum_5years,PhysRevResearch.2.043070,leonforte2024quantum,RenPRA2022}, which are two-level emitters coupled to an environment (such as a field flowing through a waveguide) at multiple spatially separated points.
As the light travels among these distinct coupling points, it accumulates a phase (optical length) $\varphi_{\text{WG}}$ proportional to the distance between them.
When the coupling points are spaced at distances comparable to the wavelength of the light they interact with, a direct consequence of the phase accumulation is that self-interference effects, absent with ordinary atoms, arise.

A remarkable feature of giant atoms is their \textit{non-Markovian }character. Indeed, a giant atom can reabsorb its own emitted excitation after a time delay proportional to the distance between the coupling points. 
This phenomenon has been experimentally demonstrated with superconducting qubits coupled to surface acoustic waves~\cite{AnderssonNatPhys2019}, and  spurred the interest in giant atoms physics.
Interestingly, as we detail below, a giant atom coupled to a waveguide at two different coupling points can be  described in terms of a  small atom (one coupling point) in a semi-infinite waveguide, Fig.~\ref{FigAnalogy}(a-b1), a typical setup to observe a non-Markovian behavior of the atomic emission~\cite{TufarelliPRA2013,TufarelliPRA2014,TudelaPRA2017}.

Another striking feature of giant atoms is the possibility of engineering dispersive decoherence-free interactions  between them~\cite{KockumPRL2018}.
Remarkably, even if the latter effect is inherently related to the phase differences associated to the displacements of the coupling points, it  becomes prominent when the travel time of light between coupling points is small compared to the characteristic time scales of the emitters, \textit{i.e.}, in the \textit{Markovian} limit~\cite{KockumPRL2018,PhysRevResearch.2.043184}.

In parallel to memory effects, another important aspect concerns 
the potential to  adjust 
the propagation direction of  light between the nodes. 
When scattered radiation displays a preferred direction, 
the interaction between  emitters and light is defined as \textit{chiral} \cite{lodahl2017chiral}.
Both theoretically \cite{RamosPRA2016} and very recently experimentally \cite{JoshiPRX2023} it has been shown that introducing light-matter couplings with an additional complex phase  can induce a chiral behaviour in the radiation emitted by giant atoms.
For a giant atom with two coupling points,  cf.~Fig.~\ref{FigAnalogy}(a), the atomic emission is chiral whenever the phase difference between the couplings $\varphi_{c}$ does not vanish.
In particular, when such a phase matches the optical length $\varphi_{\rm WG}$, 
the emission can become maximally chiral~\cite{RamosPRA2016}.

In this work, we bridge these two seemingly unrelated features of giant atoms, namely their  \textit{chirality} and their  \textit{(non-)Markovianity}.
We 
show how to make a giant atom enter the Markovian regime, even for non-negligible time delays, by tuning its chirality.
Importantly, our result depends \textit{solely} on the complex nature of the atom-light couplings.
This result definitely shows that the identifying feature of giant atoms is the non-locality of their couplings, rather than their non-Markovianity.

\textit{Setup and Hamiltonian.---}
We  consider a single giant atom weakly coupled to a one-dimensional (1D) bidirectional waveguide. The full light-matter Hamiltonian is  $\hat H = \hat H_a + \hat H_w + \hat H_\text{int}$. The free atomic Hamiltonian is  $\hat H_a =  \omega_{a}\hat \sigma ^\dagger \hat \sigma$, with 
$\hat \sigma=\dyad{g}{e}$, $\ket{g}$ and $\ket{e}$ being the ground and excited atomic states, respectively.
We model the  waveguide with a translationally invariant tight-binding array of coupled resonators with Hamiltonian
\begin{equation}\label{eq:discrWG}
	\hat H_w
	=
	-J\sum _x \hat a _{x+1}^\dagger \hat a _{x}+ \T{ H.c.}\,
\end{equation}
where $\hat a _{x}$ are  real space bosonic annihilation operators and $J>0$.
We can Fourier transform $\hat a _{x} = \sum_k e^{-ikx}  \hat a _{k}/\sqrt{N}$, where $N$ is the number of resonators, so that the Eq.~\eqref{eq:discrWG} becomes $\hat H_w = \sum_k \omega_k \hat a _{k}^\dagger \hat a _{k}$,
with $\omega_k = -2J\cos k$ (the lattice constant is set to 1).
The atom-waveguide interaction Hamiltonian is
\begin{equation}\label{eq:int_ham}
    \hat H_\T{int}
    =
    g\hat\sigma \left(e^{i\varphi_1}\hat a_{x_1}^\dagger + e^{i\varphi_2}\hat a_{x_2}^\dagger\right)+ \T{ H.c.}\,,
\end{equation}
where we assume $g$ to be real and $x_2=x_1+d$. 
We focus here on a two-legged giant atom, though our result generalizes to the the case of multiple coupling points, as we detail in~\cite{Appendix}.
In Fourier space the interaction Hamiltonian~\eqref{eq:int_ham} reads $\hat H_\T{int}
=
\sum_k 
\hat h_\T{int}(k)$,
where $\hat h_\T{int}(k) = g_k\hat\sigma\hat a_k^\dagger + \text{H.c.}$ and $g_k = g[e^{i(\varphi_1+kx_1)}+e^{i(\varphi_2+kx_2)}]/\sqrt{N}$.
When all these phases are zero, we refer to the giant atom as \textit{non-chiral}.
By contrast, we will call the giant atom \textit{chiral} whenever we take into account non-zero phases. 
This is because, in the latter case, time-reversal symmetry is broken ($g_k\neq g_{-k}$, or $T \hat h_\T{int}(k) T^{-1} \neq \hat h_\T{int}(-k)$, $T$ being the time-reversal symmetry operator).
The assumption of weak coupling makes our system equivalent to a giant atom coupled to a continuous waveguide with linear dispersion~\cite{ShenPRA2009}, see Fig.~\ref{FigAnalogy}(c). 
Therefore, from an experimental point of view, 
our waveguide Hamiltonian can be implemented with a continuous transmission line~\cite{JoshiPRX2023}, as well as with an array of coupled superconducting  circuits~\cite{KimPRX2021,ScigliuzzoPRX2022,ZhangScience2023}.

\textit{Result.---}
Our result can be condensed in the following sentence: the chirality of a giant atom  controls its Markovianity. Remarkably,  Markovianity can be achieved irrespectively of any time delay. 
This implies that such a chiral giant atom undergoes spontaneous emission  even when the coupling points are significantly far apart,  when reabsorption would occur in the non-chiral case. 
Despite the atom being \textit{giant}, in the sense that a non-Markovian behavior is expected, it behaves as if it were \textit{small} (\textit{i.e.}, single-legged). 
Thus, we argue in favor of the \textit{non-locality of the couplings} as a defining feature of giant atoms.

We derive this result through the analytic calculation of the  atomic dynamics and further check it through the Lindblad master equation. We then provide two mechanisms for this phenomenon, based on an interference argument and on a collision model picture \cite{ciccarello_quantum_2022, Ciccarello_qmetro, CilluffoPRRes2020}.

Assume the initial state is $\ket{\Psi(0)}=\ket{e}\!\ket{0}$, $\ket{0}$ being the vacuum state of the field. Then at time $t$ the full atom-waveguide state is $\ket{\Psi(t)}
	=
	\varepsilon(t)\ket{e}\!\ket{0}
	+
	\sum_k\,
	c_k(t)\,\hat a_{k}^\dagger
	\ket{g}\!\ket{0}\,.$
Imposing the Schr\"odinger equation, the dynamics of an initially excited chiral giant atom follows the delay differential equation~\cite{Appendix}
\begin{equation}\label{eq:delay}
\dot \varepsilon(t)= 
-\frac{\Gamma}{2} \varepsilon(t)
-\frac{\Gamma}{2}
\cos(\varphi_c)
e^{i\varphi_\T{WG}}
\Theta(t-t_d)
\varepsilon(t-t_d)\,.
\end{equation}
Here, $\Theta(t)$ is the Heaviside step function, $\varphi_\T{WG}$ is the optical length between the coupling points, $t_d$ is the corresponding time delay, and $\Gamma$ is the decay rate.
More specifically, the optical length is given by $\varphi_\T{WG}=k_a d$ where $d$ is the distance between the coupling points, $k_a$ is the momentum corresponding to the atomic transition frequency $\omega_{a}$ ($k_a=\pi/2$ in our case), $v=2J\sin(k_a)$ is the speed of light in the waveguide, and $\Gamma=4 g^2/v$. 

For $\varphi_c=0$, Eq.~\eqref{eq:delay} is well known~\cite{TufarelliPRA2013}, and indeed shows the analogy between a non-chiral giant atom and a small atom  in front of a mirror, cf.~Fig.~\ref{FigAnalogy}(a-b1). 
The only difference with  Ref.~\cite{TufarelliPRA2013} is the minus sign in front of the second term in the right hand side of Eq.~\eqref{eq:delay}. 
Such $\pi$ phase difference is due to the fact that for an atom in front of a mirror the optical length is proportional to twice the distance between the atom and the mirror.

\begin{figure}[t!]
	\centering
    \includegraphics[width=\columnwidth]{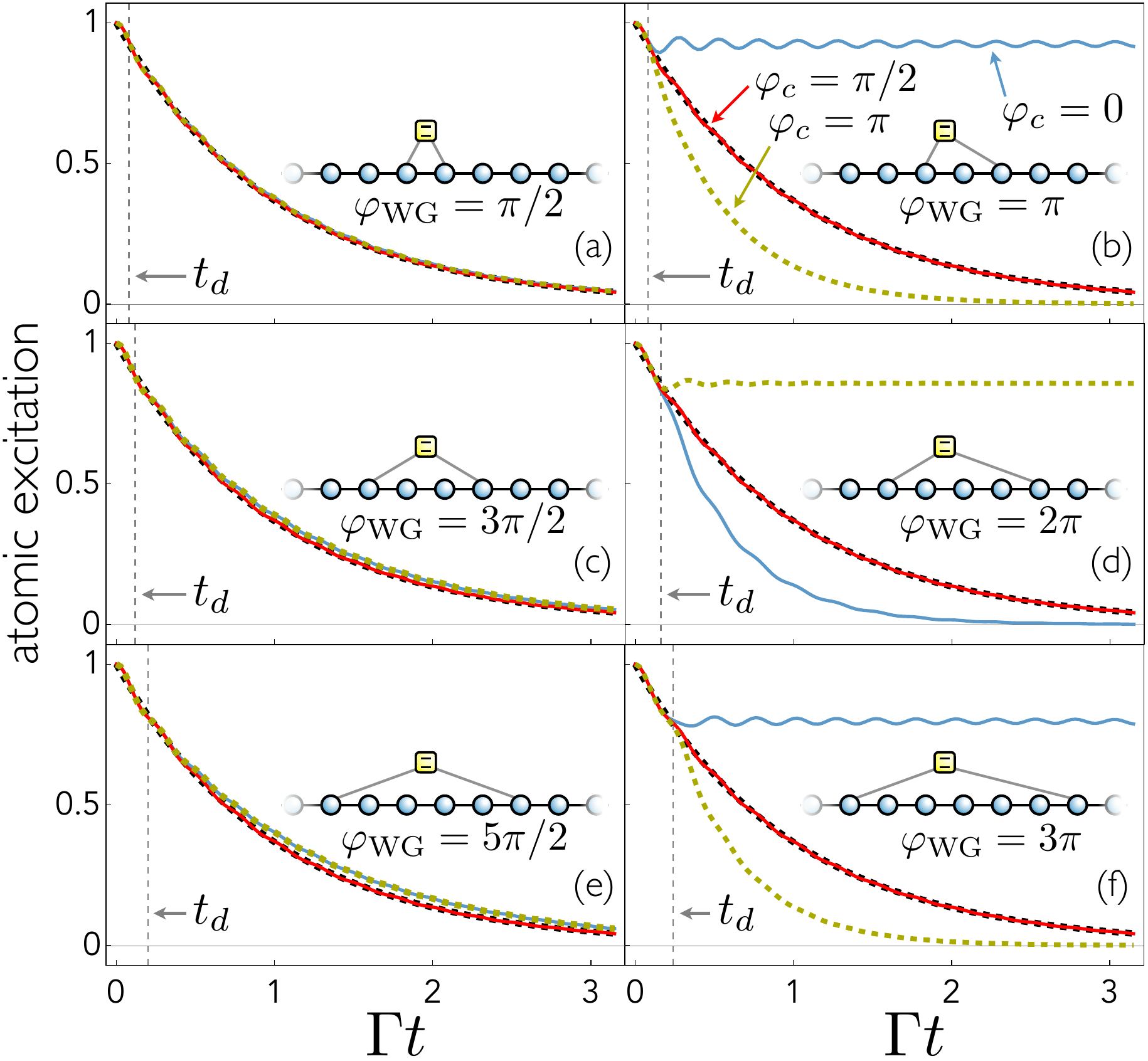}
	\caption{
		{\textit{Markovianity for any time delay.} 
		Dynamics of the atomic excitation, $|\varepsilon(t)|^2$, of an initially excited chiral giant atom for various distances $d$ between the coupling points, corresponding to various optical lengths $\varphi_\text{WG}=k_ad$ and  phase differences $\varphi_c$ between the couplings to the waveguide. 
        Specifically, $d=1,2,3,4,5,6$ in (a-f), respectively. In all panels (a-f) $\varphi_c=0,\pi/2,\pi$ correspond to blue, red and dashed yellow, respectively. The black dotted lines represent  the  exponential decay $e^{-\Gamma t}$, while the other curves are obtained numerically.
        Regardless of the time delay $t_d=d/v$, vertical dashed gray line, increasing from (a) to (f), at $\varphi_c=\pi/2$ the decay is  exactly exponential.
        When the optical length is an integer multiple of $\pi$, the atom never fully decays and correspondingly a BIC occur. On top of the BICs  occurring at odd multiples of $\pi$ for a non-chiral giant atom, (b) and (f), as in Ref.~\cite{TufarelliPRA2013},  complex couplings allow the appearance of BICs at even multiples of $\pi$ as well~\cite{Appendix}. Other parameter values: $k_a=\pi/2$, $g=0.2J$, $v=2J$, $N=90$ (number of resonators), the first (second) coupling point is at $N/2$ ($N/2+d$).
		}
	} 
	\label{fig:decayChiral}
\end{figure}

Interestingly, we observe that the atomic decay is exactly exponential at $\varphi_c=\pi/2$, regardless of the time delay $t_d$, matching the behavior of a small atom coupled  to a waveguide.
By contrast, the non-Markovianity of a chiral giant atom is prominent 
when  $\varphi_\T{WG}=m\pi$ and $\varphi_c=(m+1)\pi$  with  integer $m\geq0$.
At these values, part of the emitted light is trapped between the coupling points forming a 
bound state in the continuum (BIC)~\cite{HsuNatRevMat2016,CalajoPRL2019}.
In Fig.~\ref{fig:decayChiral}, we show the dynamics of an initially excited chiral giant atom, which indeed decays exponentially, regardless of the distance between the coupling points, for $\varphi_c=\pi/2$.

This behavior can be captured as well through the atomic master equation~\cite{Breuer2007Theory} for small distances between the coupling points.
In this case the Markov approximation, which makes  the evolution of the density matrix time local, is still valid.
Notwithstanding, the interaction Hamiltonian still keeps track of the spatial non-locality of the atom-light interaction.
The atomic master equation reads
$\dot \rho = -i[\hat H_a,\rho] + \gamma \left(\hat\sigma\rho \hat\sigma^\dagger- \{\hat\sigma^\dagger \hat\sigma, \rho\}/2 \right)$
where~\cite{Appendix} 
\begin{equation}\label{eq:ratefromME}
    \gamma = \Gamma\left[1+\cos(\varphi_c)\cos(k_a d)\right]\,.
\end{equation}
This matches the analytical result, predicting an exponential decay with rate $\Gamma$  for $\varphi_c=\pi/2$. The same rate is obtained for $k_a d=\pi/2$, which, for our discretized waveguide, corresponds to odd distances $d$. 
Indeed,  Fig.~\ref{fig:decayChiral} shows that for odd  $d$'s the atomic de-excitation slightly deviates from the exponential decay one would get from the Lindblad master equation for any complex phase $\varphi_c$. We notice though that this discrepancy increases with $d$ as the Markov approximation breaks down.

\begin{figure*}
	\centering
	\includegraphics[width=\textwidth]{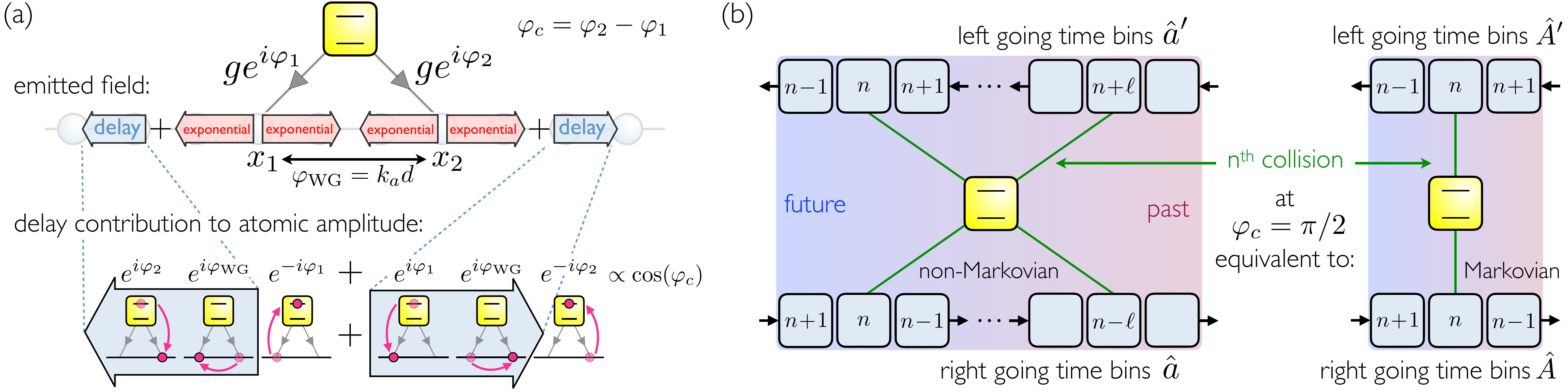}
	\caption{
		{\textit{Mechanism of Markovianity for any time delay.} 
        (a) Interference picture. At each coupling point ($x_1$, $x_2$) the emitted field can be divided into forward and backward components  (right- and left-pointing blue and red arrows, respectively) $c_{1,\T{f}}(t)$ and $c_{1,\T{b}}(t)$ (same for $x_2$). For $t\geq t_d=d/v$, $d=x_2-x_1$, the forward [backward] component at the coupling point $x_2$ [$x_1$] acquires a delay contribution $c_{2,\T{f}}^\T{del}(t)$ [$c_{1,\T{b}}^\T{del}(t)$] (light blue arrows), as the atomic excitation (small fuchsia circle) is transferred by hopping to the $x_1$ [$x_2$] coupling point and travelling along the waveguide for a distance $d$. A non-Markovian behavior takes place only if the atom gets re-excited. The amplitude for this process to occur is the sum of the two possible ways for the re-excitation to happen, 
        as illustrated at the bottom part.
        Considering the phases aquired along these two paths, the final amplitude is proportional to $\cos(\varphi_c)$. 
        (b) Collision model picture. Left- and right-going field modes (top and bottom, respectively) are mapped into trains of time-bin ancillae moving in opposite directions. At each time step the system interacts with two  separated ancillae from each bath and the chains are shift by one position. After the first collision, each ancilla will interact again with the system in $\ell$ steps. When the complex phase $\varphi_c=\pi/2$ (and odd multiples) this picture is unitarily equivalent to a Markovian collision model where the system interacts locally with the two left- and right- going time bins of a bidirectional waveguide.
        }
	} 
	\label{mech}
\end{figure*}

\textit{Mechanism.---}
First, the exponential atomic decay at $\varphi_c=\pi/2$ can be explained as an interference effect, Fig.~\ref{mech}(a).
The key observation is that the phases $\varphi_{1,2}$ and $\varphi_{\T{WG}}$ have to be considered with and without their signs, respectively. Indeed, $\varphi_{\T{WG}}$ is always positive regardless of the interference path, while $\varphi_{c}$ is positive (negative)  when going  from the atom (field) to the field (atom). 

At the left and right coupling points, $(x_1,x_2)$, we can write the emitted field amplitude at time $t$ as $c_{1,2}(t)$~\cite{Appendix}. 
We can further divide these terms into  backward (to the left) and forward (right) emitted field amplitude $c_{1}(t) = c_{1,\T{b}}(t)+c_{1,\T{f}}(t)$, analogously for $c_{2}(t)$. 
The backward  emitted field at coupling point $x_1$  has (i) a contribution coming directly from the atomic exponential decay, and (ii)
a contribution coming from the backward  emitted field at coupling point $x_2$.
Thus, we can  further divide $c_{1,\T{b}}(t)$ into its \textit{exponential} (exp) and \textit{delay} (del) contributions as  
$c_{1,\T{b}}(t)=c_{1,\T{b}}^\T{del}(t) + c_{1,\T{b}}^\T{exp}(t)$ 
and 
$c_{2,\T{f}}(t) = c_{2,\T{f}}^\T{exp}(t) + c_{2,\T{f}}^\T{del}(t)$ (the same goes for the forward emitted field at coupling point $x_2$).
Note that $c_{1,\T{b}}^\T{del}(t)=c_{2,\T{f}}^\T{del}(t)=0$ for $t<t_d$.

The only way to increase the atomic amplitude $\varepsilon$ is via the field contribution coming from the delay, which is (we drop  the time dependence to lighten notation)
\begin{equation}\label{eq:interf} 
c_{1,\T{b}}^\T{del} \, e^{-i\varphi_1}
+
c_{2,\T{f}}^\T{del} \,
e^{-i\varphi_2}\,.
\end{equation}
On the other hand, the atomic population contribution to these field components is 
\begin{equation}\label{eq:interf2}
c_{1,\T{b}}^\T{del} = e^{i\varphi_2} \, e^{i\varphi_\T{WG}} \, \varepsilon,
\qquad
c_{2,\T{f}}^\T{del} = e^{i\varphi_1} \, e^{i\varphi_\T{WG}} \, \varepsilon\,.
\end{equation} 
By plugging Eqs.~\eqref{eq:interf2} into Eq.~\eqref{eq:interf}, the delay contribution to the atomic amplitude turns out to be proportional to $\cos(\varphi_{c})$. Therefore, for $\varphi_c=\pi/2$ there is no delay contribution, and the atomic excitation decays exactly exponentially irrespectively of any distance between coupling points.

Second, 
in the framework of collision models~\cite{ciccarello_quantum_2022}, the time-evolution of the atom and the waveguide is described as a sequence of discrete interactions (collisions) involving the system (the atom) and  discretized field modes (the ancillae or time bin modes). In the case of a giant atom, the interaction involves  two separated ancillae, see Fig.~\ref{mech}(b). A non-Markovian behaviour typically arises due to the double interaction between the system and the same ancilla after a finite time.
For a two-legged giant atom, the coupling Hamiltonian \eqref{eq:int_ham} in interaction picture with respect to the waveguide and the atom reads 
\begin{eqnarray}\label{eq:Hintintpic}
\hat{H}_{\rm int}(t) 
& = & 
g\,\, \hat{\sigma}^\dagger [
e^{-i k_a x_1}\,{\hat a}_{t-\tau_1}+ e^{i k_a x_1}\,\hat {a}'_{t+\tau_1} 
\notag \\
& &
+
e^{i \varphi_c}( e^{-i k_a x_2}\,{\hat a}_{t-\tau_2}+ e^{i k_a x_2}\,{\hat a}'_{t+\tau_2})
]
\notag \\
& &
+ {\rm H.c.}
\end{eqnarray}
Without loss of generality we have set $\varphi_1=0$ and thus $\varphi_2=\varphi_c$. 
In Eq.~\eqref{eq:Hintintpic}, ${\hat a}_{t-\tau_{1,2}} ({\hat a}'_{t+\tau_{1,2}})$ are the right- (left-) going time bin operators corresponding to coupling points $x_{1,2}$~\cite{CarolloPRRes2020}, and $\tau_{1,2} = x_{1,2}/v$ are the time-domain coordinates corresponding to the coupling points' positions. 
The left-going operators have a prime to stress the distinction with the right-going ones.
For an infinitesimal evolution time $\Delta t$, the related propagator is approximated as $\mathcal{\hat{U}}_n \simeq \mathbb{1} -i ( \hat{\mathcal{H}}_n^{(0)}+\hat{\mathcal{H}}_n^{(1)})\Delta t - (\hat{\mathcal{H}}_n^{(0)})^2 \Delta t^2/2$, 
where $\hat{\mathcal{H}}_n^{(0)}=\tfrac{1}{\Delta t}\!\int_{t_{n-1}}^{t_n} \T{d}s \,\hat{H}_{\rm int}(s) $ and $\hat{\mathcal{H}}_n^{(1)} =\tfrac{i}{2\Delta t}\!\int_{t_{n-1}}^{t_n} \!\T{d}s \int_{t_{n-1}}^{s} \!\T{d}s' \,[\hat{H}_{\rm int}(s'),\hat{H}_{\rm int}(s)]$ are the $0$th and the $1$st order terms of the Magnus expansion of the generator~\cite{Magnus1954}.
The operators of both the emitter and the waveguide modes are present only in the $0$th-order term, \textit{i.e.}, only $\hat{\mathcal{H}}_n^{(0)}$ describes the interaction while $\hat{\mathcal{H}}_n^{(1)}$ is a Lamb-shift term.
We thus write the former as
\begin{align}\label{eq:interaction_collision}
\hat{\mathcal{H}}_n^{(0)} =& 
\,\,g\, \hat\sigma^\dagger 
[
 {\hat a}_n + e^{i \varphi_c} e^{-i \varphi_\T{WG}}   \,{\hat a}_{n-\ell}\,+
\notag \\&
\,\,{\hat a}_n' + e^{i \varphi_c} e^{i \varphi_\T{WG}}   \,{\hat a}_{n+\ell}'
]+ {\rm H.c.}
\end{align}
where we introduced the right-going time-bin operators $\hat a_n=\tfrac{1}{\sqrt{\Delta t}}\int_{t_{n-1}}^{t_n}\!\T{d}t\, \hat a_t$ (the same holding for the left-going ones $ \hat a_n'$). Without loss of generality, we set $x_1=0$, $x_2=d$ and $\ell \Delta t = \tau_2$ (which is nothing but the time delay $t_d$).
The relation \eqref{eq:interaction_collision} captures all the relevant physics of the atom-waveguide crosstalk in all  regimes and makes clear the (general) non-Markovian nature of the dynamics.

Consider now a separate (Markovian) collision model of an emitter coupled to a bidirectional waveguide, whose $n$th collision is described by the Hamiltonian
\begin{align}\label{eq:markovCM}
\hat{h}_n^{(0)} &= 
\sqrt{2} g\,\, \hat\sigma^\dagger 
(\hat {A}_n +  \hat {A}_n')+ {\rm H.c.}
\,,
\end{align}
where $\hat {A}_n$ and $\hat {A}_n'$ are the right- and left- going time-bin  operators, respectively.
Setting 
$\hat {A}_n = (\hat {a}_n +
e^{i(\varphi_c-\varphi_\T{WG})}
\hat {a}_{n-\ell})/\sqrt{2}$ and 
$\hat {A}_n' = (\hat {a}_n' +
e^{i(\varphi_c+\varphi_\T{WG})}
\hat {a}_{n+\ell}')/\sqrt{2}$, 
these transformations define a \textit{unitary}  transformation of the field time bins~\textit{if and only if} $\varphi_c=\pi/2$~\cite{Appendix,LoudonQTL03}.

Therefore, for $\varphi_c=\pi/2$, there
exists an exact mapping 
between Eq.~\eqref{eq:interaction_collision} (generally non-Markovian) and Eq.~\eqref{eq:markovCM}  
(Markovian). 
Under this condition, the transformation does not alter the reduced dynamics of the system.
Thus, the collision model describing a chiral giant atom with an arbitrary delay line between its legs becomes equivalent to a collision model where the atom interacts with the field at a single point with a rescaled coupling strength. 
Significantly, this implies that the reduced dynamics of the system is \textit{exactly} Markovian, even in the presence of any time delay.

\textit{Conclusion.---}
An artificial atom coupled at multiple points to a waveguide is a paradigmatic setup to observe memory effects due to self-intereference. 
We have found that this paradigm can break down when the atom-light couplings are allowed to be complex.
By properly adjusting the coupling phases, the artificial atom has an exact Markovian behavior, regardless of any inherent time delay involved in the dynamics.
This unexpected effect enriches the already exotic physics of giant atoms, opening new theoretical and experimental avenues. 
Despite most efforts are devoted to the engineering of decoherence-free Hamiltonians~\cite{leonforte2024quantum,ingelsten2024avoiding}, our results shows how unexpected phenomena can occur in the opposite regime,  that is far from protecting the atom from decoherence.
Also, the effect we find is relevant from the theoretical standpoint on its own. 
Indeed, many works righteously point out that time delays need to be neglected  to derive a master equation for  giant atoms in a waveguide~\cite{KockumPRL2018,KannanNature2020,CilluffoPRRes2020,CarolloPRRes2020}.
Our result shows that, at least for a single giant atom, there is no need to make such approximation.
We note that the  effect we described could be tested in principle by coupling  a transmon qubit either to a microwave photonic waveguide~\cite{KannanNature2020,JoshiPRX2023}, or  to an array of superconducting LC circuits~\cite{ScigliuzzoPRX2022,ZhangScience2023}.
Finally,
the connection we find between chirality and Markovianity 
could be generalized to more structured photonic environments~\cite{TudelaPRA2017,BelloSciAdv2019,LeonfortePRL2021}, which is a promising direction for future work.

\textit{Acknowledgments.---}
We thank Francesco Ciccarello, Salvatore Lorenzo and Anton F.~Kockum for useful discussions. FR thanks Chaitali Joshi, Frank Yang, and Mohammad Mirhosseini for inspiring discussions. DC thanks Thibaut Lacroix for fruitful discussions. This research was funded in part by the Luxembourg National Research Fund (FNR, Attract grant 15382998). FR acknowledges financial support from the Fulbright Research Scholar Program. DC acknowledges support from the BMBF project PhoQuant (grant no. 13N16110).

\bibliographystyle{apsrev4-1}	
\bibliography{wQEDwithChiralGiantAtomsRef}

\onecolumngrid
\appendix

\section{Derivation of the delay differential equation (3)}
\label{app:spontemTLchiralGiant}
Here we derive the exact delay differential equation governing the time evolution of the amplitude of a two-legged chiral giant atom. We consider the case of multiple legs in Section~\ref{app:Llegs}. The non-chiral case  is recovered for $\varphi_c=0$.  
Assuming the initial state is $\ket{\Psi(0)}=\ket{e}\!\ket{0}$, the state at time $t$ will be
\begin{equation}
	\ket{\Psi(t)}
	=
	\varepsilon(t)\ket{e}\!\ket{0}
	+
	\sum_k\,
	c_k(t)\,\hat a_{k}^\dagger
	\ket{g}\!\ket{0}.
\end{equation}
Observe that   the photonic amplitudes in real space are obtained as $c_x(t) = \sum_k e^{-ikx}c_k(t)/\sqrt{N}$.
Imposing the Schr\"odinger equation, and making the replacements $\varepsilon(t)\rightarrow \varepsilon(t) e^{-i\omega_{a}t}$ and $c_k(t)\rightarrow c_k(t) e^{-i\omega_{a}t}$, we get the following equations for the atomic and photonic amplitudes
\begin{eqnarray}\label{evol}
	\dot \varepsilon(t)
	& = & 
	- i \sum_k\,
	g_k^*c_k(t)\,, \label{eq:eqsatom}
	\\
	\dot c_k(t)
	& = & 
	-i(\omega_k-\omega_a) c_k(t)-i g_k\varepsilon(t)\,, \label{eq:eqsfield}
\end{eqnarray}
with initial conditions  $\varepsilon(0)=1$ and $c_k(0)=0$ for all $k$.
Integrating Eq.~\eqref{eq:eqsfield} 
and 
plugging the result back in Eq.~\eqref{eq:eqsatom} we get
\begin{equation}\label{spontemiss}
	\dot \varepsilon(t)
	=
	-
	\int_0^t\T{d}s
	\varepsilon(s)
	e^{-i\omega_{a}(s-t)}
	\sum_k
	|g_k|^2 
	e^{i\omega_{k}(s-t)}\,.
\end{equation}
Now we compute explicitly the last term in previous equation (we set $\varphi_{1}=0$ so that $\varphi_{2}=\varphi_{c}$)
\begin{eqnarray}\label{coeff}
	\sum_k
	|g_k|^2 
	e^{i\omega_{k}(s-t)}
	& = &
	\frac{g^2}{N}
	\sum_k
	\left(
	2 + e^{i(\varphi_c+kd)} + e^{-i(\varphi_c+kd)}
	\right) 
	e^{i\omega_{k}(s-t)}
	\nonumber\\
	&  &
	\text{ linearizing the waveguide dispersion because of weak coupling $\omega_k\approx v|k|$}
	\nonumber\\
	& = &
	\frac{g^2}{N}
	\sum_k
	\left(
	2 + e^{i(\varphi_c+kd)} + e^{-i(\varphi_c+kd)}
	\right) 
	e^{iv|k|(s-t)}
	\nonumber\\
	&  &
	\text{ separating positive and negative $k$'s and setting $d=vt_d$}
	\nonumber\\
	& = &
	\frac{g^2}{N}
	\sum_{k>0}
	\left(
	2 + e^{i(\varphi_c+kvt_d)} + e^{-i(\varphi_c+kvt_d)}
	\right) 
	e^{ivk(s-t)}
	\nonumber\\
	&  &
	+
	\frac{g^2}{N}
	\sum_{k<0}
	\left(
	2 + e^{i(\varphi_c+kvt_d)} + e^{-i(\varphi_c+kvt_d)}
	\right) 
	e^{-ivk(s-t)}
	\nonumber\\
	& = &
	\frac{g^2}{N}
	\sum_{k>0}
	\left(
	2 e^{ivk(s-t)}
	+ 
	e^{i\varphi_c} e^{ivk(t_d+s-t)} 
	+ 
	e^{-i\varphi_c} e^{-ivk(t_d-s+t)}
	\right) 
	\nonumber\\
	&  &
	+
	\frac{g^2}{N}
	\sum_{k<0}
	\left(
	2 e^{-ivk(s-t)}
	+ 
	e^{i\varphi_c} e^{ivk(t_d-s+t)} 
	+ 
	e^{-i\varphi_c} e^{-ivk(t_d+s-t)}
	\right) 
	\nonumber\\
	&  &
	\text{
	extending the first (second) sum, which is taken around $k_a$ ($-k_a$), to all $k$'s
	}
	\nonumber\\
	& = &
	\frac{g^2}{N}
	\sum_{k}
	\left(
	2 e^{ivk(s-t)}
	+ 
	e^{i\varphi_c} e^{ivk(t_d+s-t)} 
	+ 
	e^{-i\varphi_c} e^{-ivk(t_d-s+t)}
	\right) 
	\nonumber\\
	&  &
	+
	\frac{g^2}{N}
	\sum_{k}
	\left(
	2 e^{-ivk(s-t)}
	+ 
	e^{i\varphi_c} e^{ivk(t_d-s+t)} 
	+ 
	e^{-i\varphi_c} e^{-ivk(t_d+s-t)}
	\right) 
	\nonumber\\
	&  &
	\text{ replacing $k\rightarrow-k$ in the second sum}
	\nonumber\\
	& = &
	\frac{g^2}{N}
	\sum_{k}
	\left(
	2 e^{ivk(s-t)}
	+ 
	e^{i\varphi_c} e^{ivk(t_d+s-t)} 
	+ 
	e^{-i\varphi_c} e^{-ivk(t_d-s+t)}
	\right) 
	\nonumber\\
	&  &
	+
	\frac{g^2}{N}
	\sum_{k}
	\left(
	2 e^{ivk(s-t)}
	+ 
	e^{i\varphi_c} e^{-ivk(t_d-s+t)} 
	+ 
	e^{-i\varphi_c} e^{ivk(t_d+s-t)}
	\right) 
	\nonumber\\
	& = &
	\frac{g^2}{N}
	\sum_{k}
	\left(
	4 e^{ivk(s-t)}
	+ 
	2\cos(\varphi_c) e^{ivk(t_d+s-t)} 
	+ 
	2\cos(\varphi_c) e^{-ivk(t_d-s+t)}
	\right) 
	\nonumber\\
	& &
	\text{  setting $\Delta k=\frac{2\pi}{N}$, we get }
	\frac{1}{N}
	\sum_k
	e^{ivk(s-t)}
	=
	\frac{1}{2\pi}
	\sum_k
	e^{ivk(s-t)}\Delta k
	\rightarrow
	\int \frac{\T{d}k}{2\pi}\, e^{ivk(s-t)} = \frac{1}{v}\delta(s-t)
	\nonumber\\
	& = &
	\frac{g^2}{v}
	\left[
	4 \delta(s-t)
	+ 
	2\cos(\varphi_c) \delta(t_d+s-t) 
	+ 
	2\cos(\varphi_c) \delta(t_d-s+t) 
	\right]\,,
\end{eqnarray}
where the limit is taken as $N\rightarrow\infty$.
Finally, plugging the result back into Eq.~\eqref{spontemiss} and performing the integration (recall that $\Theta(0)=1/2$) we get Eq.~(3).

\section{Master equation of a two-legged chiral giant atom}
\label{app:meTLchiralGiant}
Here we derive the chiral giant atom decay rate in Eq.~(4) through the GKSL master equation.
The interaction Hamiltonian in interaction picture with respect to $\hat H_a + \hat H_w$ reads $\hat H_I(t) 
= e^{i\omega_a t}\hat \sigma^+  \hat B(t) +\T{ H.c.}$,
with
$\hat B(t) =
\sum_k g_k^* \hat a _ke^{-i\omega_kt}$.
Notice that the information on the non-locality of the interaction is within $g_k$.
Following the standard recipe~\cite{Breuer2007Theory}, the giant atom master equation reads as $\dot \rho = -i[\hat H_a,\rho] + \gamma \left(\hat\sigma\rho \hat\sigma^\dagger- \{\hat\sigma^\dagger \hat\sigma, \rho\}/2 \right)$, where $\gamma=2\Re[I(\omega_a)]$ and 
\begin{eqnarray}
    I(\omega_a)
    =
    \int_0^\infty
    \T{d}\tau
    \langle
    \hat B(\tau)\hat B^\dagger(0)
    \rangle
    e^{i\omega_a\tau}
    & \rightarrow &
    2g^2\int_0^\infty
    \T{d}\tau
    \int \frac{\T{d}k}{2\pi} [1+\cos(\varphi_c+kd)] e^{i(\omega_a-\omega_k)\tau}
    \nonumber\\
    & = &
    2g^2\int \frac{\T{d}k}{2\pi} [1+\cos(\varphi_c+kd)] 
    \left[
    \pi\delta(\omega_a-\omega_k)+i\mathcal P \frac{1}{\omega_a-\omega_k}
    \right]\,,
\end{eqnarray}
where the waveguide state is assumed to be the vacuum, the limit is taken as $N\rightarrow\infty$, and $\mathcal P$ denotes the principal value.
Therefore, analogously to the steps taken to get Eq.~\eqref{coeff}, we get
\begin{eqnarray}
    \gamma
    & = &
    \frac{2g^2}{v}\int \T{d}k [1+\cos(\varphi_c+kd)]\delta(k_a-|k|)
    \nonumber\\
    & = &
    \frac{2g^2}{v}\int_{\{-k_a\}} \T{d}k [1+\cos(\varphi_c+kd)]\delta(k_a+k)
    +
    \frac{2g^2}{v}\int_{\{k_a\}} \T{d}k [1+\cos(\varphi_c+kd)]\delta(k_a-k)
    \nonumber\\
    & = &
    \frac{2g^2}{v}\int \T{d}k [1+\cos(\varphi_c+kd)]\delta(k_a+k)
    +
    \frac{2g^2}{v}\int \T{d}k [1+\cos(\varphi_c+kd)]\delta(k_a-k)
    \nonumber\\
    & = &
    \frac{2g^2}{v}[1+\cos(\varphi_c-k_ad)]
    +
    \frac{2g^2}{v}[1+\cos(\varphi_c+k_ad)]
    \nonumber\\
    & = &
    \frac{4g^2}{v}[1+\cos(\varphi_c)\cos(k_ad)]\,.
    \nonumber\\
\end{eqnarray}
On the one hand this shows that for $\varphi_c=\pi/2$ we get the exponential decay with the correct rate, independently on the distance between the coupling points. On the other hand, we get the same rate when $k_ad$ is an odd multiple of $\pi/2$. For small $d$ (\textit{i.e.}, when the Markovian approximation is more accurate), this matches the \textit{almost} exponential decay we see in Fig.~2 for odd $d$'s (regardless of the coupling phase).

\section{Emitted field by a chiral two-legged giant atom}\label{app:emittField}

Here we solve Eq.~\eqref{eq:eqsfield} for the field amplitudes. 
The output of this calculation makes clear the difference between the  exponential and the delay components of the field amplitudes used in the main text.
Observe that
\begin{eqnarray}\label{eq:fieldampl}
	c_x(t) 
	= 
	\frac{1}{\sqrt{N}}
	\sum_k e^{-ikx}c_k(t)
	=
	-i
	\int_0^t\T{d}s\,\varepsilon(s)
	e^{-i\omega_{a}(s-t)}
	\frac{1}{\sqrt{N}}
	\sum_k 
	g_ke^{-ikx}
	e^{i\omega_k(s-t)}\,.
\end{eqnarray}
We now compute explicitly the last term (we keep $\varphi_{1}$), analogously to the derivation of Eq.~\eqref{coeff}:
\begin{eqnarray}
	\frac{1}{\sqrt{N}}
	\sum_k 
	g_ke^{-ikx}
	e^{i\omega_k(s-t)}
	& = & 
	\frac{ge^{i\varphi_1}}{N}
	\sum_k 
	\left(
	e^{ik(x_1-x)}
	+
	e^{i\varphi_c}
	e^{ik(x_1+vt_d-x)}
	\right)
	e^{iv|k|(s-t)}
	\nonumber\\
	& = & 
	\frac{ge^{i\varphi_1}}{N}
	\sum_{k>0} 
	\left(
	e^{ik(x_1-x)}
	+
	e^{i\varphi_c}
	e^{ik(x_1+vt_d-x)}
	\right)
	e^{ivk(s-t)}
	\nonumber\\
	& & 
	+
	\frac{ge^{i\varphi_1}}{N}
	\sum_{k<0} 
	\left(
	e^{ik(x_1-x)}
	+
	e^{i\varphi_c}
	e^{ik(x_1+vt_d-x)}
	\right)
	e^{-ivk(s-t)}
	\nonumber\\
	& = & 
	\frac{ge^{i\varphi_1}}{N}
	\sum_{k} 
	\left(
	e^{ik[(x_1-x)+v(s-t)]}
	+
	e^{i\varphi_c}
	e^{ik[(x_1-x)+v(t_d+s-t)]}
	\right)
	\nonumber\\
	& & 
	+
	\frac{ge^{i\varphi_1}}{N}
	\sum_{k} 
	\left(
	e^{ik[(x_1-x)-v(s-t)]}
	+
	e^{i\varphi_c}
	e^{ik[(x_1-x)-v(-t_d+s-t)]}
	\right)
	\nonumber\\
	& = & 
	\frac{ge^{i\varphi_1}}{N}
	\sum_{k} 
	\left(
	e^{ik[(x_1-x)+v(s-t)]}
	+
	e^{i\varphi_c}
	e^{ik[(x_1-x)+v(t_d+s-t)]}
	\right)
	\nonumber\\
	& & 
	+
	\frac{ge^{i\varphi_1}}{N}
	\sum_{k} 
	\left(
	e^{ik[v(s-t)-(x_1-x)]}
	+
	e^{i\varphi_c}
	e^{ik[v(s-t-t_d)-(x_1-x)]}
	\right)
	\nonumber\\
	& \rightarrow & 
	ge^{i\varphi_1}
	\int \frac{\T{d}k}{2\pi}
	\left(
	e^{ik[(x_1-x)+v(s-t)]}
	+
	e^{i\varphi_c}
	e^{ik[(x_1-x)+v(t_d+s-t)]}
	\right)
	\nonumber\\
	& & 
	+
	ge^{i\varphi_1}
	\int \frac{\T{d}k}{2\pi}
	\left(
	e^{ik[v(s-t)-(x_1-x)]}
	+
	e^{i\varphi_c}
	e^{ik[v(s-t-t_d)-(x_1-x)]}
	\right)
	\nonumber\\
	& = & 
	\frac{ge^{i\varphi_1}}{v}
	\left[
	\delta\left(s-t+\frac{x_1-x}{v}\right)
	+
	e^{i\varphi_c}
	\delta\left(s-t+\frac{x_1-x}{v}+t_d\right)
	\right]
	\nonumber\\
	&  & 
	+
	\frac{ge^{i\varphi_1}}{v}
	\left[
	\delta\left(s-t-\frac{x_1-x}{v}\right)
	+
	e^{i\varphi_c}
	\delta\left(s-t-\frac{x_1-x}{v}-t_d\right)
	\right]\,.
	\nonumber\\
\end{eqnarray}

We plug this back into Eq.~\eqref{eq:fieldampl}, and integrating we get

\begin{eqnarray}
	c_x(t)
	& = &
	-\frac{ige^{i\varphi_1}}{v}
	\left[
	\Theta\left(t-\frac{x_1-x}{v}\right)
	\Theta\left(\frac{x_1-x}{v}\right)
	\varepsilon\left(t-\frac{x_1-x}{v}\right)
	e^{ik_{a}(x_1-x)}
	\right]
	\nonumber\\
	&  &
	-\frac{ige^{i\varphi_1}}{v}
	e^{i\varphi_c}
	\left[
	\Theta\left(t-\frac{x_1-x}{v}-t_d\right)
	\Theta\left(\frac{x_1-x}{v}+t_d\right)
	\varepsilon\left(t-\frac{x_1-x}{v}-t_d\right)
	e^{ik_{a}(x_1-x+vt_d)}
	\right]
	\nonumber\\
	&  &
	-\frac{ige^{i\varphi_1}}{v}
	\left[
	\Theta\left(t+\frac{x_1-x}{v}\right)
	\Theta\left(-\frac{x_1-x}{v}\right)
	\varepsilon\left(t+\frac{x_1-x}{v}\right)
	e^{-ik_{a}(x_1-x)}
	\right]
	\nonumber\\
	&  &
	-\frac{ige^{i\varphi_1}}{v}
	e^{i\varphi_c}
	\left[
	\Theta\left(t+\frac{x_1-x}{v}+t_d\right)
	\Theta\left(-\frac{x_1-x}{v}-t_d\right)
	\varepsilon\left(t+\frac{x_1-x}{v}+t_d\right)
	e^{-ik_{a}(x_1-x+vt_d)}
	\right]\,.
	\nonumber\\
\end{eqnarray}
In the previous equation, the first two lines correspond to the forward modes while the other two lines to the backward modes,
so that
\begin{equation}
	c_x(t) =  c_{x,\T{f}}(t) + c_{x,\T{b}}(t)\,.
\end{equation}
At the coupling points $x=x_1$ and $x=x_1+vt_d\equiv x_2$
we get ($c_{x_j}(t)\equiv c_{j}(t)$)
\begin{eqnarray}
	c_1(t) 
	& = &
	c_{1,\T{b}}(t) + c_{1,\T{f}}(t)\,,
	\nonumber\\
	c_2(t) 
	& = &
	c_{2,\T{b}}(t) + c_{2,\T{f}}(t)\,,
	\nonumber\\
\end{eqnarray}
where 
\begin{eqnarray}
	c_{1,\T{b}}(t) 
	& = &
	-\frac{ige^{i\varphi_1}}{v}
	\left[
	\varepsilon(t)
	+
	2 e^{i(\varphi_\T{WG}+\varphi_c)} \Theta(t-t_d)\varepsilon(t-t_d)
	\right]
	\equiv c_{1,\T{b}}^\T{exp}(t) + c_{1,\T{b}}^\T{del}(t)\,,
	\nonumber\\
	c_{l,\T{f}}(t) 
	& = &
	-\frac{ige^{i\varphi_1}}{v}
	\varepsilon(t)
	\equiv
	c_{1,\T{f}}^\T{exp}(t)\,,
	\nonumber\\
	c_{r,\T{b}}(t) 
	& = &
	-\frac{ige^{i\varphi_1}}{v}
	e^{i\varphi_c}
	\varepsilon(t)
	\equiv c_{2,\T{b}}^\T{exp}(t)\,,
	\nonumber\\
	c_{2,\T{f}}(t) 
	& = &
	-\frac{ige^{i\varphi_1}}{v}
	\left[
	e^{i\varphi_c}
	\varepsilon(t)
	+
	2 e^{i\varphi_\T{WG}} \Theta(t-t_d)\varepsilon(t-t_d)
	\right]
	\equiv c_{2,\T{f}}^\T{exp}(t) + c_{2,\T{f}}^\T{del}(t)\,.
	\nonumber\\
\end{eqnarray}
From these equations we see that the delay contributions are only in the forward and backward emitted field at coupling point $x_2$ and $x_1$, respectively.

\section{Bound states in the continuum (BICs)}\label{app:bics}
Here we find the conditions under which BICs occur for a two-legged chiral giant atom.
Consider a general atom-waveguide state
\begin{equation}\label{stazstate}
	\ket{\Psi}
	=
	\varepsilon\ket{e}\!\ket{0}
	+
	\sum_{x=-\infty}^\infty\,
	c_x\,\hat a_{x}^\dagger
	\ket{g}\!\ket{0}\,.
\end{equation}
Imposing the stationary Schr\"odinger equation $\hat H\ket{\Psi} = E_b \ket{\Psi}$ and then normalizability we get
\begin{eqnarray}
	E_b\varepsilon
	& = &
	\omega_a \varepsilon
	+
	g \sum_{x} c_x\,\left(e^{-i\varphi_1}\delta_{1x} + e^{-i\varphi_2}\delta_{2x}\right)\,,
	\nonumber\\
	E_b c_x
	& = &
	-J\left(c_{x-1}+c_{x+1}\right)
	+
	g\varepsilon\left(e^{i\varphi_1}\delta_{1x} + e^{i\varphi_2}\delta_{2x}\right)\,,
	\nonumber
\end{eqnarray}
where $\delta_{1x}\equiv \delta_{x_1x}$ (same for $\delta_{2x}$).
Imposing that the BIC is resonant with the bare atomic state, \textit{i.e.}~$E_b=\omega_a$, we get
\begin{eqnarray}
	0
	& = &
	e^{-i\varphi_1}c_1 + e^{-i\varphi_2}c_2\,,\label{eigenstatesdiscrete}
	\\
	vk_a c_x
	& = &
	-J\left(c_{x-1}+c_{x+1}\right)
	+
	g\varepsilon\left(e^{i\varphi_1}\delta_{1x} + e^{i\varphi_2}\delta_{2x}\right)\,.\label{eigenstatesdiscrete2}
\end{eqnarray}
Eq.~\eqref{eigenstatesdiscrete2} can be solved by Fourier transform (which amounts to make the following replacements:
$c_n \rightarrow c_k$, $c_{n\pm1} \rightarrow e^{\mp ik}c_k$, $\delta_{1x} \rightarrow  e^{ikx_1}/\sqrt{N}$, analogously for $\delta_{2x}$), getting:
\begin{eqnarray}
	c_k
	& = &
	\frac{g\varepsilon}{\sqrt{N}}\frac{e^{i(\varphi_1+kx_1)} + e^{i(\varphi_2+kx_2)}}{vk_a+2J\cos k}
	\nonumber\\
\end{eqnarray}
Back to real space, with steps analogous to those in the derivation of Eq.~\eqref{coeff}:
\begin{eqnarray}
	c_x
	& = &
	\frac{g\varepsilon}{N}\sum_k \frac{e^{i(\varphi_1+k(x_1-x))} + e^{i(\varphi_2+k(x_2-x))}}{vk_a+2J\cos k}
	\nonumber\\
	& = &
	\frac{g\varepsilon}{N}\sum_{k>0} \frac{e^{i(\varphi_1+k(x_1-x))} + e^{i(\varphi_2+k(x_2-x))}}{vk_a+2J\cos k}
	+
	\frac{g\varepsilon}{N}\sum_{k<0} \frac{e^{i(\varphi_1+k(x_1-x))} + e^{i(\varphi_2+k(x_2-x))}}{vk_a+2J\cos k}
	\nonumber\\
	& = &
	\frac{g\varepsilon}{vN}\sum_{k>0} \frac{e^{i(\varphi_1+k(x_1-x))} + e^{i(\varphi_2+k(x_2-x))}}{k_a+k}
	+
	\frac{g\varepsilon}{vN}\sum_{k<0} \frac{e^{i(\varphi_1+k(x_1-x))} + e^{i(\varphi_2+k(x_2-x))}}{k_a-k}
	\nonumber\\
	& = &
	\frac{g\varepsilon}{vN}\sum_{k} \frac{e^{i(\varphi_1+k(x_1-x))} + e^{i(\varphi_2+k(x_2-x))}}{k_a+k}
	+
	\frac{g\varepsilon}{vN}\sum_{k} \frac{e^{i(\varphi_1+k(x_1-x))} + e^{i(\varphi_2+k(x_2-x))}}{k_a-k}
	\nonumber\\
	& \rightarrow &
	\frac{g\varepsilon}{v}\int \frac{\T{d}k}{2\pi} \frac{e^{i(\varphi_1+k(x_1-x))} + e^{i(\varphi_2+k(x_2-x))}}{k_a+k}
	+
	\frac{g\varepsilon}{v}\int \frac{\T{d}k}{2\pi} \frac{e^{i(\varphi_1+k(x_1-x))} + e^{i(\varphi_2+k(x_2-x))}}{k_a-k}
	\nonumber\\
	& = &
	-\frac{g\varepsilon}{v}
	\left(
	e^{i\varphi_1}\sin\left[k_a(x_1-x)\right] + e^{i\varphi_2}\sin\left[k_a(x_2-x)\right]
	\right)
	\nonumber\\
	&  &
	+ \frac{2g\varepsilon}{v}
	\left(
	e^{i\varphi_1}\sin\left[k_a(x_1-x)\right]\Theta(x_1-x) + e^{i\varphi_2}\sin\left[k_a(x_2-x)\right]\Theta(x_2-x)
	\right)\,.
\end{eqnarray}
Plugging this back into  Eqs.~\eqref{eigenstatesdiscrete} we get the condition
\begin{equation}\label{firstcond}
	\cos(\varphi_c)\sin(\varphi_{\T{WG}})=0\,,
\end{equation}
for the occurrence of BICs.
If $\varphi_c=\pi/2$, then the state in Eq.~\eqref{stazstate} would not be normalizable. Indeed
\begin{equation}
	|c_x|^2
	=
	\frac{g^2|\varepsilon|^2}{v^2}
	\left(
	\sin^2[k_a(x_1-x)]+ \sin^2[k_a(x_2-x)]
	\right)\,,
	\quad\T{ for } x>x_2 \T{ and } x<x_1\,.
\end{equation}
Since we are looking for normalizable states, it must be $\varphi_c\neq\pi/2$,  implying $\varphi_{\T{WG}}=m\pi$.
Imposing the latter, we get $c_1=c_2=0$ and
\begin{equation}
	|c_x|^2
	=
	\frac{2g^2|\varepsilon|^2}{v^2}
	\left[
	1+(-1)^m\cos(\varphi_c)
	\right]
	\sin^2[k_a(x_1-x)]
	\,,
	\quad\T{ for } x<x_1 \T{ and } x>x_2 \,.
\end{equation}
This can only be zero if $\varphi_c=(m+1)\pi$,  yielding $c_x=0$ for $x\leq x_1$ and $x\geq x_2$ and 
\begin{equation}
	|c_x|^2
	=
	\frac{4g^2|\varepsilon|^2}{v^2}
	\sin^2\left[k_a(x_1-x)\right]
	\,,
	\quad\T{ for } x_1<x<x_2\,,
\end{equation}
where $d=m\pi/k_0$, $d=x_2-x_1$.
The normalizability of the state implies
$|\varepsilon|^2=1/\left(1+\Gamma t_d/2\right)$ which is exactly the asymptotic value of the atomic amplitude (when finite). We observe how this value does not depend on the chirality of the giant atom and matches the result in Ref.~\cite{TufarelliPRA2013}.

To summarize, given a momentum $k_a$, a BIC exists if $d=m\pi/k_a$ and $\varphi_c=(m+1)\pi$ and reads (up to an irrelevant global phase)
\begin{equation}
	\ket{\Psi}
	=
	\frac{1}{\sqrt{1
			+
			\frac{\Gamma t_d}{2}}}
	\left(
	\ket{e}
	+
	\frac{2ge^{i\varphi_1}}{v}
	\sum_{x=x_1}^{x_1+d}
	\sin[k_a(x-x_1)]
	\right)\,.
\end{equation}
We can conclude that: without a complex coupling, as in Ref.~\cite{TufarelliPRA2013}, a BIC only exist for $d$ equals odd multiples of $\pi/k_0$; with a complex coupling with $\pi$ phase a BIC only exist for $d$ equals even multiples of $\pi/k_0$. For any other phase of the complex coupling a BIC does not exist.

\subsubsection{Asymptotic atomic amplitude}\label{asintatomappendix}
The asymptotic value of the atomic amplitude can be computed as well via the Laplace transform as we do here for completeness.
Applying the Laplace transform, \textit{i.e.}~$\tilde \varepsilon(s)=\int_{0}^\infty\T{d}t\,e^{-st}\varepsilon(t)$, to Eq.~(4) amounts to making the following replacements: $\varepsilon(t)\rightarrow\tilde \varepsilon(s)$, 
$\dot \varepsilon(t)\rightarrow s\tilde \varepsilon(s)-\varepsilon(0)$, and
$\Theta(t-t_d)\varepsilon(t-t_d)\rightarrow e^{-st_d}\tilde \varepsilon(s)$, so to get 
\begin{equation}
	\tilde \varepsilon(s)
	=
	\frac{1}{s+\frac{\Gamma}{2}\left[1+e^{i\varphi_{\T{WG}}}
		e^{-st_d}\cos(\varphi_c)\right]}\,.
\end{equation}
Using the property that $\varepsilon(t\rightarrow\infty)=\lim_{s\rightarrow0}s\tilde \varepsilon(s)$ 
we get
\begin{equation}\label{asinteq}
	\varepsilon(t\rightarrow\infty)
	=
	\left(1+\frac{\Gamma t_d}{2}\right)^{-1}
	\delta(\varphi_{\T{WG}}-m\pi)
	\delta(\varphi_c-(m+1)\pi)\,.
\end{equation}
This is nothing but the result obtained for an atom in front of a mirror~\cite{TufarelliPRA2013}, except that in that case, instead of  two $\delta$ functions, we would only have $\delta(\varphi_{\T{WG}}-2(m+1)\pi)$.

\section{Unitary nonlocal-to-local mapping for a two-legged atom}
\label{app:mechanism2leg}
\begin{figure*}
	\centering
	\includegraphics[width=0.85\textwidth]{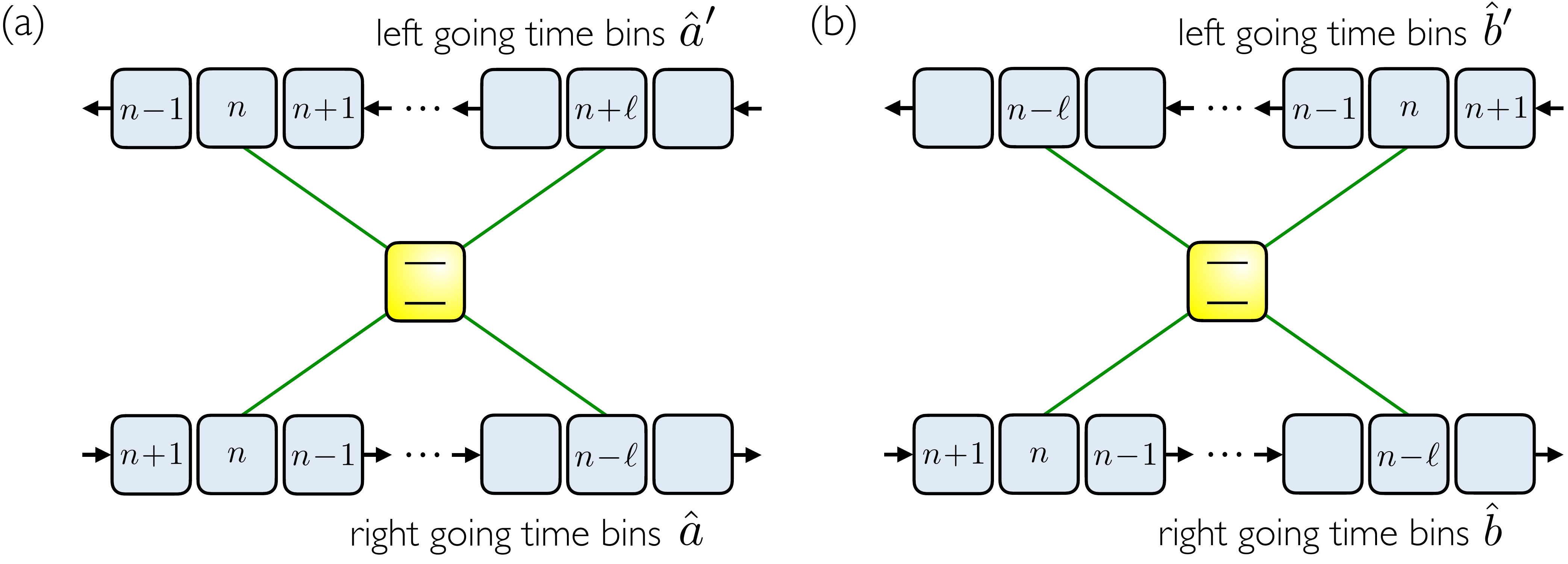}
	\caption{
		\textit{Alternative labeling conventions for bidirectional field time-bins.} 
		(a) The time bins interacting with the atom at the left coupling points are both labeled as ``$n$''. This is the convention used in Fig.3(b).
		(b) In both baths, $n$ labels the two fresh time bins colliding with the system for the first time.
	} 
	\label{relab}
\end{figure*}

The atom-waveguide interaction term in the case of two coupling points is given by Eq.~(9). We report it here again for the sake of clarity
\begin{align}
	\hat{\mathcal{H}}_n^{(0)} =& 
	\,\,g\, \hat\sigma^\dagger 
	[
	{\hat a}_n + e^{i \varphi_c} e^{-i \varphi_\T{WG}}   \,{\hat a}_{n-\ell}\,+
	\,\,{\hat a}_n' + e^{i \varphi_c} e^{i \varphi_\T{WG}}   \,{\hat a}_{n+\ell}'
	]+ {\rm H.c.}
\end{align}
Note that the primed (left-going mode) and unprimed (right-going mode) operators have different subscripts, reflecting the fact that they are traveling in opposite directions, with the same convention used in Ref.~\cite{CilluffoPRRes2020}. Equivalently, we can adopt the same labeling convention of right-going modes for left-going modes, as depicted in Fig.~\ref{relab}, by defining a new family of operators $\hat{b}_m$ such that $\hat{b}_m = \hat{a}_m$ and $\hat{b}'_m = \hat{a}'_{m+l}$.
Thus we consider the couple of time-bin ladders $ v= (\hat{b}_{n},\hat{b}_{n-l})$ under the condition that the first element corresponds to the right-going bath, while the second pertains to the left-going bath ($\hat{b}_{n-l}' = \hat{a}_{n}$).
It is easy to find that the composite bath operators in Eq.~(9) can be generated through the action of a global transformation over $v$ as follows
\begin{align}
	&\frac{1}{\sqrt{2}}
	\begin{pmatrix}
		1 & e^{i \varphi_c} e^{-i \varphi_\T{WG}}  \\
		e^{i \varphi_c} e^{i \varphi_\T{WG}} & 1 
	\end{pmatrix}
	\begin{pmatrix}
		\hat {b}_{n} \\
		\hat {b}_{n-l} 
	\end{pmatrix}
	= \frac{1}{\sqrt{2}}
	\begin{pmatrix}
		{\hat b}_{n} + e^{i \varphi_c} e^{-i \varphi_\T{WG}}\,\hat {b}_{n-l} \\
		e^{i \varphi_c} e^{i \varphi_\T{WG}}\, \hat{b}_{n} +  \hat {b}_{n-l}
	\end{pmatrix}\,
	= \frac{1}{\sqrt{2}}
	\begin{pmatrix}
		{\hat a}_{n} + e^{i \varphi_c} e^{-i \varphi_\T{WG}}\,{\hat a}_{n-l} \\
		e^{i \varphi_c} e^{i \varphi_\T{WG}}\, \hat{a}'_{n+l} +  \hat {a}'_{n}
	\end{pmatrix}\,.
\end{align}
The transfer matrix is unitary if and only if $\varphi_c = \pi/2$ (and odd multiples). 
Given that the open dynamics is invariant under global unitary transformations over the environment (\textit{i.e.}~the waveguide), we can deduce that under this condition, the collision model where the atom interacts with the field through two points with an arbitrary delay, is equivalent to a collision model where the atom locally couples to two independent baths. Therefore the dynamics is Markovian even in the presence of arbitrarily large time delays.
Note that this operation is possible because of the presence of two baths. In the context of a chiral waveguide, an extension of the bath could be identified to meet the same condition. This generally nontrivial operation exemplifies a case of Markovian embedding.

\section{Collision model description of chiral emission in the Markovian limit}
Here, we note  how the collision model picture provides a simple explanation of the chirality of the emission given by a complex coupling~\cite{RamosPRA2016}. In the case $\varphi_\T{WG}= \pi/2$ (necessary for maximal chiral emission) and in the negligible-time-delay limit  $(\ell \sim 0)$, Eq.(8) in the main text reduces to
\begin{equation*}
\hat{\mathcal{H}}_n^{(0)} = 
g\, \hat{\sigma}^\dagger [
 (1-ie^{i \varphi_c}) {\hat a}_{n} + 
 (1+ie^{i \varphi_c})\,{\hat a}_{n}' ]
+ {\rm H.c.}
\end{equation*}
Hence, adjusting the complex phase $\varphi_c$ enables control over the chirality of the emission. Specifically, when $\varphi_c=\pi/2$, emission to the left-going modes is inhibited, 
in agreement with the results in Refs.~\cite{RamosPRA2016,JoshiPRX2023}.

\section{Generalization to multiple coupling points}
\label{app:Llegs}

Here we show that a Markovian behavior can be obtain even with multiple coupling points, by properly tuning the coupling phases.
The atom-light interaction Hamiltonian now reads
\begin{equation}\label{eq:int_hamapp}
	\hat H_\T{int}
	=
	g\hat\sigma \left(\sum_{j=1}^{L} e^{i\varphi_j}\hat a_{x_j}^\dagger\right)
	+ \T{ H.c.}\,,
\end{equation}
Without loss of generality we can set $x_1=0$.
We notice that Eq.~\eqref{spontemiss} is valid for any number of coupling points. Thus, we need to compute only its last term (as we did in Appendix~\ref{app:spontemTLchiralGiant}):
\begin{eqnarray}\label{coeff3L}
&&\sum_k
|g_k|^2 
e^{i\omega_{k}(s-t)}=
\nonumber\\
& = &
\frac{g^2}{N}
\sum_k
\left( L + \sum_{j=2}^L e^{i[\varphi_j+k(j-1)d]} + \sum_{j=2}^L e^{-i[\varphi_j+k(j-1)d]} + \sum_{\substack{j,n=2\\ j\neq n}}^L e^{i[(\varphi_j-\varphi_n)+k(j-n)d]} \right)
e^{i\omega_{k}(s-t)}
\nonumber\\
& = &
\frac{g^2}{N}
\sum_k
\left( L + \sum_{j=2}^L e^{i[\varphi_j+k(j-1)d]} + \sum_{j=2}^L e^{-i[\varphi_j+k(j-1)d]} + \sum_{\substack{j,n=2\\ j\neq n}}^L e^{i[(\varphi_j-\varphi_n)+k(j-n)d]} \right)
e^{iv|k|(s-t)}
\nonumber\\
& = &
\frac{g^2}{N}
\sum_{k>0}
\left( L + \sum_{j=2}^L e^{i[\varphi_j+(j-1)kvt_d]} + \sum_{j=2}^L e^{-i[\varphi_j+(j-1)kvt_d]} + \sum_{\substack{j,n=2\\ j\neq n}}^L e^{i[(\varphi_j-\varphi_n)+(j-n)kvt_d]} \right)
e^{ivk(s-t)}
\nonumber\\
&  &
+
\frac{g^2}{N}
\sum_{k<0}
\left( L + \sum_{j=2}^L e^{i[\varphi_j+(j-1)kvt_d]} + \sum_{j=2}^L e^{-i[\varphi_j+(j-1)kvt_d]} + \sum_{\substack{j,n=2\\ j\neq n}}^L e^{i[(\varphi_j-\varphi_n)+(j-n)kvt_d]} \right)
e^{-ivk(s-t)}
\nonumber\\
& = &
\frac{g^2}{N}
\sum_{k}
\left( 
L e^{ivk(s-t)}
+ 
\sum_{j=2}^L e^{i\varphi_j} e^{ivk[(j-1)t_d+s-t]} 
+ 
\sum_{j=2}^L e^{-i\varphi_j} e^{-ivk[(j-1)t_d-s+t]}
+ 
\sum_{\substack{j,n=2\\ j\neq n}}^L
e^{i(\varphi_j-\varphi_n)}
e^{ivk[(j-n)t_d+s-t]} 
\right)
+
\nonumber\\
&  &
\frac{g^2}{N}
\sum_{k}
\left( 
L e^{-ivk(s-t)}
+ 
\sum_{j=2}^L e^{i\varphi_j} e^{ivk[(j-1)t_d -s+t]} 
+ 
\sum_{j=2}^L e^{-i\varphi_j} e^{-ivk[(j-1)t_d +s-t]} 
+ 
\sum_{\substack{j,n=2\\ j\neq n}}^L e^{i(\varphi_j-\varphi_n)}
e^{ivk[(j-n)t_d-s+t]} 
\right)
\nonumber\\
& = &
\frac{g^2}{N}
\sum_{k}
\left( 
L e^{ivk(s-t)}
+ 
\sum_{j=2}^L e^{i\varphi_j} e^{ivk[(j-1)t_d+s-t]} 
+ 
\sum_{j=2}^L e^{-i\varphi_j} e^{-ivk[(j-1)t_d-s+t]}
+ 
\sum_{\substack{j,n=2\\ j\neq n}}^L
e^{i(\varphi_j-\varphi_n)}
e^{ivk[(j-n)t_d+s-t]} 
\right)
+
\nonumber\\
&  &
\frac{g^2}{N}
\sum_{k}
\left( 
L e^{ivk(s-t)}
+ 
\sum_{j=2}^L e^{i\varphi_j} e^{-ivk[(j-1)t_d -s+t]} 
+ 
\sum_{j=2}^L e^{-i\varphi_j} e^{ivk[(j-1)t_d +s-t]} 
+ 
\sum_{\substack{j,n=2\\ j\neq n}}^L e^{i(\varphi_j-\varphi_n)}
e^{-ivk[(j-n)t_d-s+t]} 
\right)
\nonumber\\
& = &
\frac{g^2}{N}
\sum_{k}
\left( 
2 L e^{ivk(s-t)}
+ 
2\sum_{j=2}^L \cos(\varphi_j) e^{ivk[(j-1)t_d+s-t]} 
+ 
2 \sum_{j=2}^L \cos(\varphi_j) e^{-ivk[(j-1)t_d-s+t]}
\right)
\nonumber\\
&  &
+
\frac{g^2}{N}
\sum_{k}
\left( 
2 \sum_{\substack{j,n=2\\ j > n}}^L
\cos(\varphi_j-\varphi_n)
e^{ivk[(j-n)t_d+s-t]} 
+ 
2 \sum_{\substack{j,n=2\\ j < n}}^L
\cos(\varphi_n-\varphi_j)
e^{-ivk[(n-j)t_d-s+t]}
\right)
\nonumber\\
& \rightarrow &
\frac{g^2}{v}
\left( 
2 L \delta(s-t)
+ 
2\sum_{j=2}^L \cos(\varphi_j) \delta((j-1)t_d+s-t) 
+ 
2 \sum_{j=2}^L \cos(\varphi_j) \delta((j-1)t_d-s+t)
\right)+
\nonumber\\
& &
\frac{g^2}{v}
\left( 
2 \sum_{\substack{j,n=2\\ j > n}}^L
\cos(\varphi_j-\varphi_n)
\delta((j-n)t_d+s-t) 
+ 
2 \sum_{\substack{j,n=2\\ j < n}}^L
\cos(\varphi_n-\varphi_j)
\delta((n-j)t_d-s+t)
\right)\,.
\end{eqnarray}
Plugging back into  Eq.~\eqref{spontemiss} we get
\begin{eqnarray}\label{eq:spontemNL}
\dot \varepsilon(t)
& = &
-
\int_0^t\T{d}s
\varepsilon(s)
e^{-i\omega_{a}(s-t)}
\sum_k
|g_k|^2 
e^{i\omega_{k}(s-t)}
\nonumber\\
& \rightarrow &
-\frac{g^2}{v}
\int_0^t\T{d}s
\varepsilon(s)
e^{-ivk_{a}(s-t)}
\times
\nonumber\\
&  &
\times
\left( 
2 L \delta(s-t)
+ 
2\sum_{j=2}^L \cos(\varphi_j) \delta((j-1)t_d+s-t) 
+
2 \sum_{\substack{j,n=2\\ j > n}}^L
\cos(\varphi_j-\varphi_n)
\delta((j-n)t_d+s-t) 
\right)
\nonumber\\
& = &
-\frac{g^2}{v}
\left( 
L \varepsilon(t)
+
2\sum_{j=2}^L \cos(\varphi_j)
e^{ivk_{a}(j-1)t_d}
\Theta(t-(j-1)t_d)
\varepsilon(t-(j-1)t_d)
\right)
\nonumber\\
&  &
-\frac{g^2}{v}
\left( 
2 \sum_{\substack{j,n=2\\ j > n}}^L
\cos(\varphi_j-\varphi_n)
e^{ivk_{a}(j-n)t_d}
\Theta(t-(j-n)t_d)
\varepsilon(t-(j-n)t_d)
\right)
\nonumber\\
& = &
-\frac{g^2}{v}
\left( 
L \varepsilon(t)
+
2\sum_{j=2}^L \cos(\varphi_j)
e^{i(j-1)\varphi_\text{WG}}
\Theta(t-(j-1)t_d)
\varepsilon(t-(j-1)t_d)
\right)
\nonumber\\
&  &
-\frac{g^2}{v}
\left( 
2 \sum_{\substack{j,n=2\\ j > n}}^L
\cos(\varphi_j-\varphi_n)
e^{i(j-n)\varphi_\text{WG}}
\Theta(t-(j-n)t_d)
\varepsilon(t-(j-n)t_d)
\right).
\end{eqnarray}
Now with algebraic manipulation we get (we restored $\varphi_1$ to unify the notation, though it can still be safely set to zero)
\begin{equation}
    \dot \varepsilon(t)
    =
    -\frac{g^2}{v}
    \left( 
    L \varepsilon(t)
    +
    2
    \sum_{n=1}^{L-1}
    \left[
    \left(
    \sum_{j=1}^{L-n}
    \cos(\varphi_{n+j}-\varphi_j)
    \right)
    e^{in\varphi_\text{WG}}
    \Theta(t-nt_d)
    \varepsilon(t-nt_d)
    \right]
    \right)
\end{equation}
Therefore a Markovian behavior is achieved when the phases $\varphi_2,\ldots,\varphi_L$
satisfy the following $L-1$ equations
\begin{equation}
    \sum_{j=1}^{L-n}
    \cos(\varphi_{n+j}-\varphi_j)=0
\end{equation}
for $n=1,\ldots,L-1$.
When these equations are satisfied, the decay rate of the giant atom is $Lg^2/v$, namely $L$ times the rate of a small atom in a waveguide.
A general solution to these coupled equations is not informative on its own, rather we notice that one can always solve the last one  ($n=L-1$), finding $\varphi_L=\pi/2$, plugging it back in the second to last one ($n=L-2$), and recursively find a solution.
For instance, with three coupling points ($L=3$), a Markovian bahavior is achieved by setting $\varphi_3=\pi/2$ and $\varphi_2=-\pi/4$, while with four coupling points ($L=4$) one can set $\varphi_4=\varphi_2=\pi/2$ and $\varphi_3=\pi$.

\end{document}